# Characterization of Pd and Pd@Au core-shell nanoparticles using atom probe tomography and field evaporation simulation

Se-Ho Kim[a,b], Kyuseon Jang[a], Phil Woong Kang[c], Jae-Pyoung Ahn[d], Jae-Bok Seol[e], Chang-Min Kwak[f], Constantinos Hatzoglou[g], François Vurpillot[g], Pyuck-Pa Choi[a*]

[a]Department of Materials Science and Engineering, Korea Advanced Institute of Science and Technology (KAIST) 291 Daehak-ro, Yuseong-gu, Daejeon 34141, Republic of Korea

[b]Max-Planck-Institut für Eisenforschung GmbH, Max-Planck-Straße 1, 40237, Düsseldorf, Germany

[c]Department of Chemical and Biomolecular Engineering, Korea Advanced Institute of Science and Technology (KAIST) 291 Daehak-ro, Yuseong-gu, Daejeon 34141, Republic of Korea

[d]Advanced Analysis Center, Korea Institute of Science and Technology (KIST), Seoul 136-791, Republic of Korea

[e]National Institute for Nanomaterials Technology (NINT), POSTECH, Pohang 790-784, Republic of Korea

[f]Department of Materials Science and Engineering, POSTECH, Pohang 790-784, South Korea

[g]Normandie Université, UNIROUEN, INSA Rouen, CNRS, Groupe de Physique des Matériaux, 76000 Rouen, France

*Corresponding author

Email: p.choi@kaist.ac.kr

**Abstract**

We report on atom probe tomography analyses of Pd and Pd@Au nanoparticles embedded in a Ni matrix and the effects of local evaporation field variations on the atom probe data. In order to assess the integrity of the reconstructed atom maps, we performed numerical simulations of the field evaporation processes and compared the simulated datasets with experimentally acquired data. The distortions seen in the atom maps for both Pd and Pd@Au nanoparticles could be mostly ascribed to local variations in chemical composition and elemental evaporation fields. The evaporation field values for Pd and Ni, taken from the image hump model and assumed in the simulations, yielded a good agreement between experimental and simulation results. In contrast, the evaporation field for Au, as predicted from the image hump model, appeared to be substantially overestimated and resulted in a large discrepancy between experiments and simulations.

**Introduction**

Nanoparticles with controlled sizes and shapes have received much attention due to their outstanding electronic [1–3], optical [4–6], magnetic [7–9], and catalytic [10–12] properties. These properties are usually strongly influenced by the chemical composition and elemental distribution and hence, a key to understanding the structure-property relationships of nanoparticles [13–16] is their characterization down to the atomic scale. Except for a very few cases [17–20] this goal has not been achieved yet, as most of the characterization techniques available for nanoparticles do not show sufficient chemical sensitivity and/or spatial resolution.

Atom probe tomography (APT) is a powerful technique that could overcome the limitations of conventional characterization methods. It combines the capability of performing three-dimensional analyses with near atomic spatial resolution and ppm-level detection sensitivity [21–23]. The fundamental physical process on which APT relies is field evaporation [24]. A high positive DC voltage applied to a sharp needle-shaped specimen together with ultra-short voltage or laser pulses induces controlled field desorption of surface atoms. Desorbed atoms are ionized and accelerated towards a position sensitive detector, where the positively charged ions follow the electric field lines of the tip. Chemical information is obtained from time-of-flight mass spectrometry. By applying an inverse projection algorithm, the positions (x, y, and z coordinates) of the detected ions can be reconstructed from the detector impact positions and the detection sequence [25,26].

Thus, APT is in principle an ideal tool for mapping the elemental distribution within nanoparticles with high accuracy. However, the fabrication of APT samples from nanoparticles poses a great challenge as nanoparticles must be somehow incorporated into needle-shaped specimens of about 100 nm in thickness. This is why reports about successful APT measurements on nanoparticles are still limited [27–34].

Recently, we proposed a new method of preparing APT specimens from Pd nanoparticles, which is based on depositing nanoparticles on a metal substrate using electrophoresis followed by electrodeposition of a Ni film to embed the nanoparticles [35]. Needle-shaped APT specimens could be routinely prepared from the Pd nanoparticle/Ni composite film using focused ion beam (FIB) milling [36]. In this work, we exploit this approach for performing comparative APT analyses and numerical simulations of the field evaporation process on well-defined Pd and Pd@Au core-shell nanoparticles. In order to assess the APT experiments and the reconstructed APT data, simulated datasets were acquired using a software developed by Vurpillot et al. [37]. Based on a two-dimensional cell model for the needle-shaped specimen, the electric field, ion trajectories and impact coordinates on a virtual detector were calculated, where the latter were used for a two-dimensional reconstruction of the simulated APT data. The aim of this paper is to discuss the aberrations in elemental distributions due to local field variations by comparing atom maps reconstructed from experimental and simulated data.

**Materials and Methods**

**Synthesis of Pd and Pd@Au nanoparticles**

All chemical reagents used for nanoparticle synthesis were purchased from Sigma-Aldrich, namely sodium tetrachloropalladate (II) ($Na_2PdCl_4$, 98 %), gold (III) chloride trihydrate ($HAuCl_4 \cdot 3H_2O$, 99.9 %), L-ascorbic acid (AA, 99 %), poly-vinylpyrrolidone (PVP-55, MW=55 000), potassium bromide (KBr, 99 %), and hexadecyltrimethylammonium bromide (CTAB, MW=364.45). Nickel sulfate heptahydrate ($NiSO_4 \cdot 6H_2O$, Junsei Chemical Co.), nickel chloride hexahydrate ($NiCl_2 \cdot 6H_2O$, Samchun Chemical Co.), and boric acid ($H_3BO_3$, Samchun Chemical Co.) were used for the electroplating process. All aqueous solutions were prepared using deionized (DI) water.

Pd nanoparticles were synthesized according to Ref. [38] by mixing 8 ml of DI water, 300 mg of KBr, 60 mg of L-ascorbic acid, and 105 mg of poly-vinylpyrrolidone (PVP) and dissolving 57 mg of the Pd precursor ($Na_2PdCl_4$) into the solution at a temperature of about 80 ºC. After 3 h, the precipitated nanoparticles were extracted using a centrifuge and rinsed several times with DI water to remove remnants of the chemical solutions.

Pd@Au nanoparticles were synthesized according to Ref. [39]. 30 mg of the Au precursor ($HAuCl_4$), 5 mg of PVP, and 6 mg of L-ascorbic acid were dissolved in 8 ml of DI water in a three-necked flask containing a condenser and a stir bar. The flask was heated to 95 ºC and an as-synthesized Pd nanoparticle solution was slowly added under vigorous stirring for 12 h, resulting in the formation of core-shell Pd@Au nanoparticles in solution. After cooling to room temperature the solution was washed with DI water several times and re-dispersed in 10 mL of DI water.

**Electrophoresis and Ni electroplating**

Pd and Pd@Au nanoparticles were embedded in a Ni film, using a sequential two-step process involving electrophoresis and electrodeposition [35]. The zeta potential of each synthesized nanoparticle solution was measured using a zeta potential analyzer (Malvern, Zetasizer nano-zs). 0.01 M of CTAB in DI water was added to increase the zeta potential up to +30 mV and ensure colloidal stability and successful electrophoresis. The nanoparticle solution was poured into a vertical cell containing a Cu substrate at the bottom. A constant current of -10 mA was applied for 100 s to deposit the nanoparticles on the Cu substrate. Subsequently, the solution for electrophoresis was replaced by a Watts solution [40] to electrodeposit a Ni film. A constant current of -100 mA was applied for 200 s, which resulted in a film thickness of about 6 μm.

**TEM and APT characterization**

As-synthesized and embedded nanoparticles were characterized using transmission electron microscopy (TEM) (JEOL JEM-2100F) in scanning TEM-high angle annular dark field (STEM-HAADF) mode. APT specimens were prepared from the electrodeposited nanoparticle/Ni films using a FIB instrument (Helios Nanolab 450) and standard FIB lift-out protocols [36] (see Fig. S1 for an example of a sharpened APT tip). A final annular milling step was performed at 5 kV to position nanoparticles close to the apex and to reduce Ga implantation.

A Cameca LEAP 4000X HR instrument was used to perform APT analyses in pulsed laser mode. Specimen base temperatures of 60 K and 100 K, laser pulse energies of 80 pJ and 100 pJ, and pulse frequencies of 125 kHz and 100 kHz were chosen for Pd and Pd@Au nanoparticle specimens, respectively, at a detection rate of 0.5 % for both cases. The latter were prone to premature fracture during APT runs, and thus a relatively high base temperature and laser energy were selected. Data analyses were done using the commercial software, Imago Visualization and Analysis Software (IVAS) 3.6.14, developed by Cameca Instruments. Three-dimensional atom maps were reconstructed using the standard protocol developed by Bas et al. [25] and Geiser et al. [26].

**Simulation of the field evaporation process**

In order to assess the effects of chemical variations and nanoparticle shapes on the reconstructed APT data, we performed numerical simulations of the field evaporation of a Pd (or Pd@Au) nanoparticle in a Ni matrix using a software developed by Vurpillot et al. [41]. This software has been recently implemented into in the 3.8.2 version of IVAS. It enables the calculation of the gradual evolution of the sample under field evaporation, assuming a sample with rotational symmetry along the z axis. Phases of interests are included inside the sample, assuming constant field evaporation for each phase. Schematic drawings of the initial simulation conditions are illustrated in Fig. 1 for both a Pd and Pd@Au nanoparticle embedded in the Ni matrix. We chose a radius of 50 nm and a shank angle of 20º for the simulated tips, which represents a realistic geometry for actual APT samples. The nanoparticle sizes in the simulations (Pd (10 nm) and Pd@Au (10@5 nm)) corresponded to the average sizes and shapes of the actual as-synthesized nanoparticles.

For simplicity, we assumed a simple rigid square lattice for the simulations, where each square-shaped cell had a length of 0.3 nm and was considered as a single atom with a defined evaporation field. Temperature effects were neglected, i.e. a temperature of 0 K was assumed [42,43]. The electrostatic potentials of all points in the vacuum between the tip surface and the detector were calculated for a standing voltage ($V = 1$) applied to the sample with a geometrically defined standard local electrode in front of the sample. Both the local electrode and the detector were considered to be electrostatically grounded ($V = 0$). The voltage distribution was calculated after the field evaporation of each surface atom by numerically solving the Laplace equation using an iterative finite difference algorithm. Once the equi-potentials were calculated, the electric field, $F$, above all surface atoms was computed. Within a single phase, the cell submitted to the highest electric field was chosen as the atom to be field evaporated. For heterogeneous samples composed of atoms $i$ with various evaporation fields $F_i$, the probability of field evaporation was defined as $\omega_i = F/F_i$. Thus, maximum $\omega_i$ indicated the atom that would be field evaporated next. Once this process occurred, the geometrical cell of the field evaporated atom was transformed into a vacuum cell and the electric field distribution was recalculated. By computing the Newton's equation of motion, the trajectory of the ion from the initial atom position to the defined virtual detector was numerically computed. This process was re-iterated for several thousands of atoms evaporated one by one from the tip apex. The impact positions of the field evaporated ions on the virtual detector were used to reconstruct a two-dimensional map from the simulated data.

For each field evaporated atom, the local field above this atom before evaporation was equal to the assumed evaporation field, $F_i$. The actual voltage $V^*$ necessary to reach this effective field $F_i$ was estimated to be $V^* = F_i/F$ using a simple rule of thumb. In this way, ion impacts on the virtual detector, and voltage variation on the sample, were numerically computed. Further details on the numeric simulation of the field evaporation process may be found in Ref. [41,43–47].

In the first simulation test we used evaporation field values of Pd (37 V nm$^{-1}$), Au (53 V nm$^{-1}$), and Ni (35 V nm$^{-1}$) derived from the image hump model (Ref. [48]) to define the evaporation field ratios between the nanoparticle and the matrix as well as between the nanoparticle core and its shell (see Table 1). Subsequently, the evaporation field of Au was varied to assess the difference between the three-dimensional atom maps reconstructed from the experimental and simulated data.

**Results**

**TEM characterization of as-synthesized Pd and Pd@Au nanoparticles**

As-synthesized Pd nanoparticles showed cuboidal shape and an average size of about 10 nm, as confirmed by TEM investigations (see Fig. 2a). The cuboidal Pd nanoparticles were also used as seeds for synthesizing the Pd@Au core-shell nanoparticles. As shown in the STEM-HAADF image and STEM-EDS map in Fig. 2b, the average size of Pd@Au core-shell nanoparticles was about 20 nm. The Au shells were grown spherically on the surface of the cuboidal Pd nanoparticles and consequently the thickness of the Au shells varied from 5 to 10 nm. Furthermore, both the as-synthesized Pd nanoparticles and Pd cores of the core-shell Pd@Au nanoparticles exhibited slightly rounded corners.

**APT analyses of Pd and Pd@Au nanoparticles embedded in Ni**

Fig. 3a shows a three-dimensional atom map of a Pd nanoparticle in the Ni matrix reconstructed using the standard reconstruction protocol proposed by Bas et al. [25] and Geiser et al. [26]. Using an iso-concentration surface of 50 at. % Pd (purple) the Pd nanoparticle could be highlighted and its shape was found to be approximately cuboidal and its size about 10 nm in reasonable agreement with TEM observations. As visualized by the iso-concentration surface, the reconstructed nanoparticle shape exhibited slightly rounded corners. We note that the particle was adjacent to another partially detected particle at its right bottom corner. Furthermore, a thin slice of the atom map (see Fig. 3b) revealed crystal lattice planes, indicating that the crystallinity of the nanoparticle was well-preserved during its synthesis and the subsequent APT sample preparation.

In contrast to Pd, the reconstructed atom map of a Pd@Au nanoparticle (see Fig. 4a) showed strong deviations from its original shapes. Both the corners of the reconstructed Pd core and the Au shell were distorted and exhibited spikes (see Fig. 4b). These spikes are related to rapidly changing evaporation fields at the Pd/Au and Au/Ni interfaces, as demonstrated in the field evaporation simulations below. Fig. 4c shows an atomic density profile and elemental concentration profiles through the Pd@Au core-shell nanoparticle along the z-direction. Two peaks could be observed in the density profile at the Ni/Au and Au/Ni interfaces. Furthermore, strong intermixing between the Ni matrix atoms and Au shell atoms could be observed, where the apparent Au concentration in the shell was only about 25 at. % at maximum. The apparent intermixing between Ni and Au could be ascribed to reconstruction artifacts, which will be discussed in detail below. Moreover, concentration fluctuations appeared in the Au concentration profile, which originated from the spikes in the atom map (see Fig. 4a).

**Simulation of the field evaporation process**

Fig. 5a shows the map of a Pd nanoparticle (purple) in the Ni matrix (green) reconstructed from a simulated dataset. The relative field between particle and matrix was set to 1.1, according to the evaporation field values for Pd (37 V $nm^{-1}$) and Ni (35 V $nm^{-1}$) predicted from the image hump model [48]. The reconstructed Pd particle exhibited nearly cuboidal shape in reasonable agreement with the experimentally acquired data (see Fig. 3a). However, its top and bottom edge showed slightly reduced and increased atomic density, respectively, due to changing evaporation fields at the horizontal Pd/Ni interfaces.

Fig. 5b shows the map of a Pd@Au nanoparticle in Ni reconstructed from a simulated dataset. In this first simulation test, an evaporation field value of 53 V $nm^{-1}$ was used for Au according to the image hump model [48] and the same values for Pd and Ni as above. Green, blue, and purple dots represent the reconstructed positions of Ni, Au, and Pd atoms, respectively. The Au shell appeared severely distorted and exhibited several spikes at the interface with the Ni matrix. However, one-dimensional concentration and density profiles plotted along the z-direction across the Pd/Au/Ni interfaces showed strong discrepancies to the experimental data in Fig. 4c. More specifically, no intermixing could be observed between the Ni matrix and the Au shell. Due to this lack of agreement between the experimental and simulated data, we iteratively reduced the evaporation field of the Au shell and repeated the simulations.

Fig. 6a-c show the simulated map of a Pd@Au nanoparticle embedded in Ni with different Au evaporation field (35, 28, and 25 V nm$^{-1}$). Significant intermixing between Ni and Au at the horizontal Ni/Au interfaces and in particular the spikes in the Au distribution, as seen in the experimental data (compare Fig. 4b), could only be observed for an assumed evaporation field 25 V nm$^{-1}$ for Au (Fig. 6c). The corresponding 1D concentration and density profiles across the Pd@Au nanoparticle in Ni with different evaporation field of Au are shown in Fig. 7a-d. The closest agreement between experiments and simulations is obtained for Fig. 7d (25 V nm$^{-1}$), as expected from the atom maps.

**Discussion**

Both Pd and Au are noble metals with high standard electrode potentials, and hence galvanic corrosion is negligible during the embedding process of the nanoparticles in Ni by means of electrodeposition. Thus, it is of key interest to elaborate on what may have caused the shape distortions in the three-dimensional atom maps for both the Pd and the Pd@Au nanoparticles. Artifacts, i.e. errors in calculating the spatial atomic positions, generally prevail in APT data reconstructed with the standard point projection protocol [25,26] and can be usually ascribed to different evaporation fields of different elements. It has been reported by several authors that local variations in evaporation field in heterogeneous materials, in particular at chemical interfaces, strongly affect the integrity of the reconstructed APT data [43,49,50]. For instance, density fluctuations at layer interfaces were reported for thin-film multilayers consisting of a low-field material and a high-field material [43,47]. Such density fluctuations match those observed for the top and bottom interfaces of the cuboidal Pd nanoparticles with the Ni matrix (see Fig. 5a). Furthermore, a local protrusion may develop at the corner of the Pd nanoparticle, as Pd shows a slightly higher evaporation field than Ni. Such a protrusion is expected to cause a local magnification effect [51] and the observed rounded corners observed in the atom maps. However, this effect seems small and difficult to assess as Pd nanoparticles already showed slightly rounded corners in the as-synthesized state. Overall, the assumed evaporation fields of Ni (35 V $nm^{-1}$) and Pd (37 V $nm^{-1}$), as predicted by the image hump model [24], yield simulation results in reasonable agreement with the experiments.

In contrast, the evaporation field of Au (53 V $nm^{-1}$) predicted by the image hump model, resulted in strong discrepancies between the reconstructed simulation and experimental data. As opposed to the simulations, the experimental data revealed intermixing between Ni and Au. Thus, we performed further simulations assuming various evaporation fields of Au (53, 35, 28, and 25 V $nm^{-1}$) and examined the corresponding changes in the reconstructed APT data. Fig. 7a-d show density and concentration profiles plotted vertically along the reconstructed Pd@Au core-shell nanoparticle for the various cases.

The closest agreement between simulated and experimental data could be obtained for an evaporation field 25 V $nm^{-1}$ for Au. The intermixing at the Ni/Au interfaces seen in the profiles and atom maps as well as the Au spikes could only be reproduced for this value in the simulated data. These spikes observed at the Au/Ni and Au/Pd interfaces in the 3D reconstructions may be ascribed to rapid changes in local evaporation rate, whenever Au atoms are just exposed to the surface, and the associated shape distortions of the tip surface. Overall, the comparative studies of this work indicate that the actual evaporation of Au is about 25 V nm$^{-1}$ and hence substantially lower than the theoretical evaporation field (53 V nm$^{-1}$) predicted by the image hump model.

As seen in Fig. 3a and 5a, the reconstructed atom maps of Pd in Ni showed only little distortion, since both Pd and Ni were found to have nearly identical evaporation fields. Therefore, the occurrence of the Au spikes could be partly mitigated by using other metallic matrices such as Sn (23 V nm$^{-1}$) or Zr (28 V nm$^{-1}$) that have similar evaporation fields to Au as we determined (25 V nm$^{-1}$). However, reconstruction artifacts at the Au/Pd interface are an intrinsic problem of Pd@Au core-shell nanoparticles, which cannot be avoided with the standard reconstruction protocol by Bas et al. [25] and Geiser et al. [26]. More advanced reconstruction algorithms, taking into account variations in local evaporation fields and tip radii, are needed [52–54].

**Conclusions**

Pd and Pd@Au nanoparticles embedded in a Ni matrix were investigated by APT measurements and numerical simulations of the field evaporation process. Slight and strong deviations from the original as-synthesized shapes could be observed for the reconstructed APT datasets of Pd and Pd@Au nanoparticle samples, respectively. The reconstruction artifacts could be ascribed to the difference in evaporation fields of Pd, Au, and Ni. In particular, the large difference in evaporation field between Au and Ni (or Pd) was held responsible for the strongly distorted reconstructed shapes of Pd@Au nanoparticles. An assumed value of 53 V nm$^{-1}$ for Au, as predicted from the image hump model, yielded strong discrepancies between experimental and simulation results and much better agreement was achieved by assuming a substantially lower evaporation field of 25 V nm$^{-1}$ for Au.

**Acknowledgement**

This research was supported by the National Research Foundation of Korea (NRF) (grant number: 2019R1A2C1002165). S.-H. K acknowledges financial support from the ERC-CoG-SHINE-771602 and J.B.S. acknowledges financial support from the NRF (grant number: 2018R1C1B6008585).

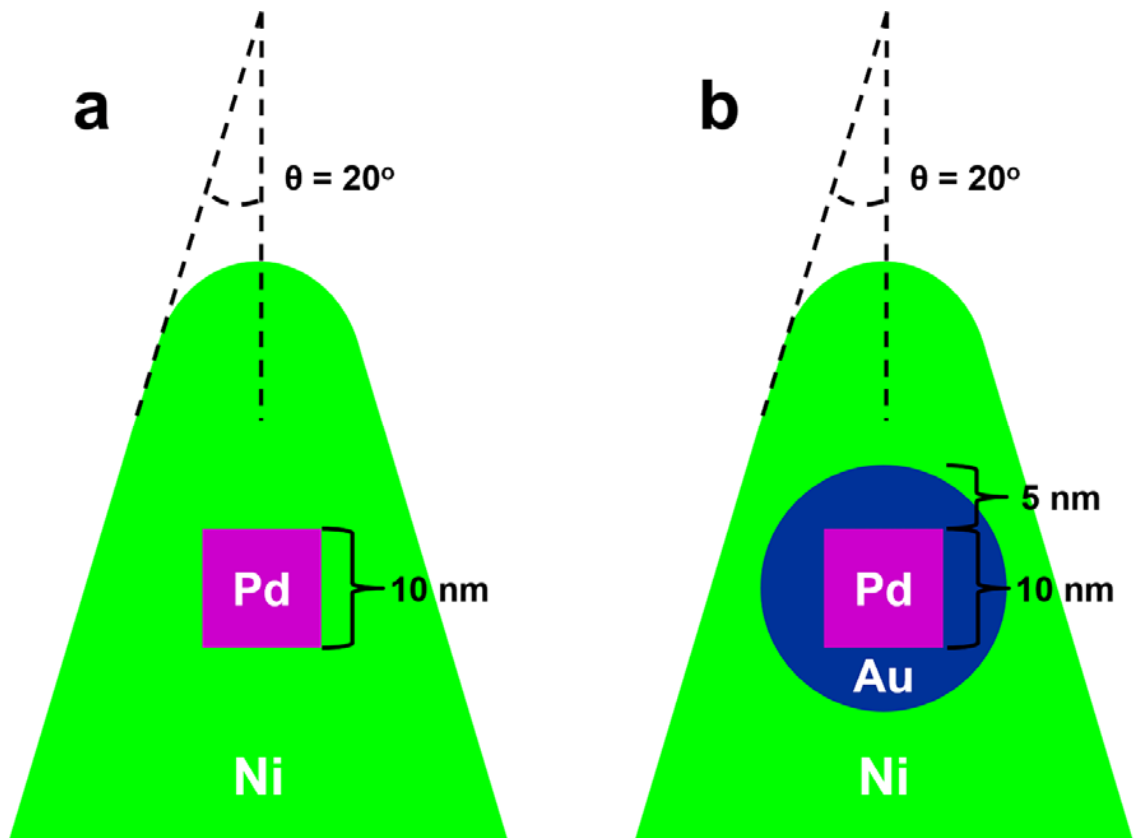

**Fig. 1.** Schematics of the initial conditions for numerical simulation of the field evaporation process of (a) square Pd nanoparticle (purple) and (b) Pd@Au nanoparticle with a square core (purple) and a circular shell (blue) within a Ni matrix. The initial radius of curvature of the specimen was set to 50 nm.

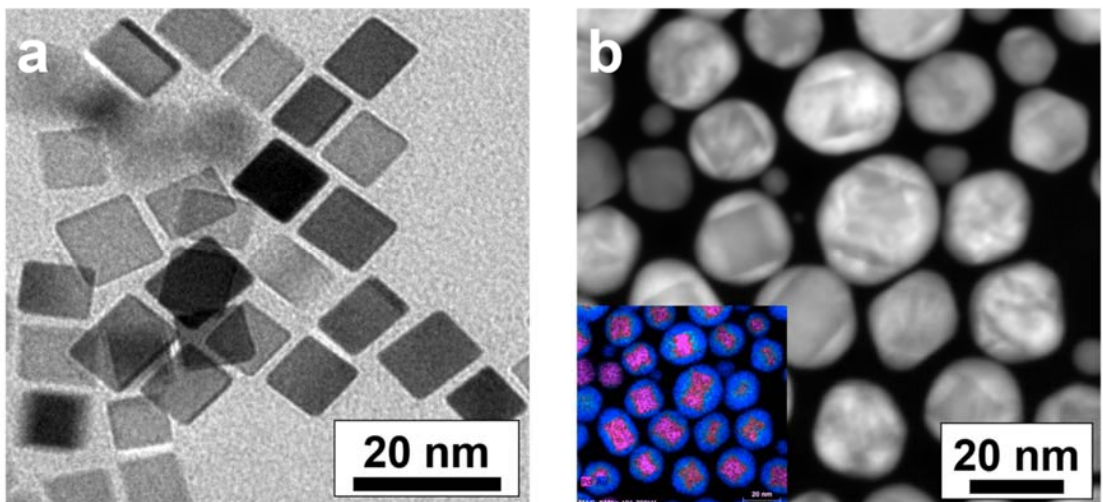

**Fig. 2.** (a) TEM and (b) HAADF-STEM images of as-synthesized Pd nanoparticles and Pd@Au core-shell nanoparticles, respectively. The inset shows an EDS map of as-synthesized Pd@Au nanoparticles: Pd (purple) and Au (blue).

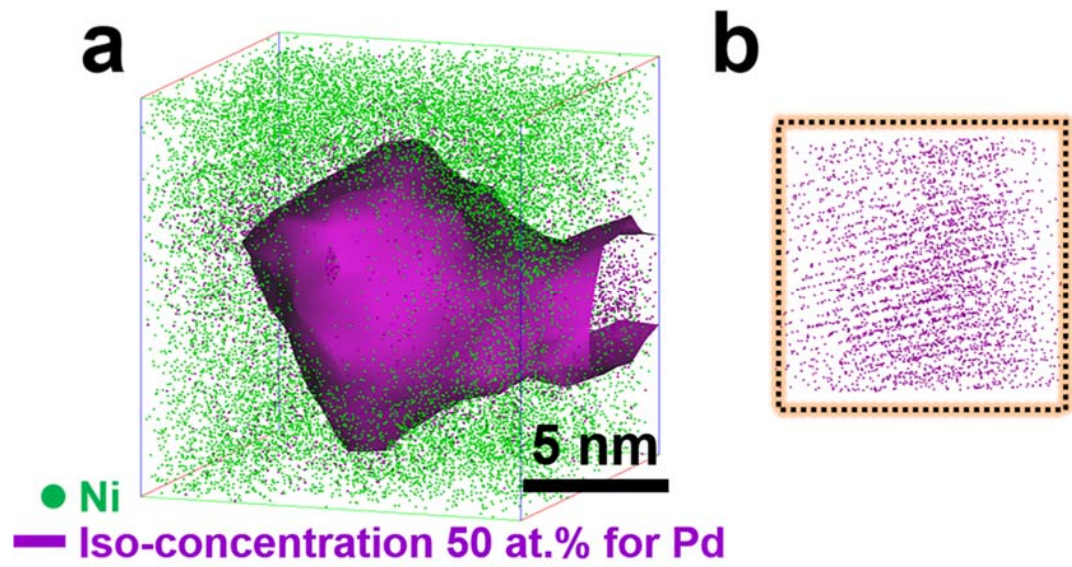

**Fig. 3.** (a) Reconstructed 3D atom map of Pd nanoparticles embedded in Ni. Green and purple dots mark the reconstructed positions of Ni and Pd, respectively. Two detected Pd nanoparticles are highlighted by 50 at. % Pd iso-concentration surfaces. The bottom right edge of a Pd nanoparticle overlaps with another one. (b) Lattice planes of Pd in the sectioned volume from Fig. 3a.

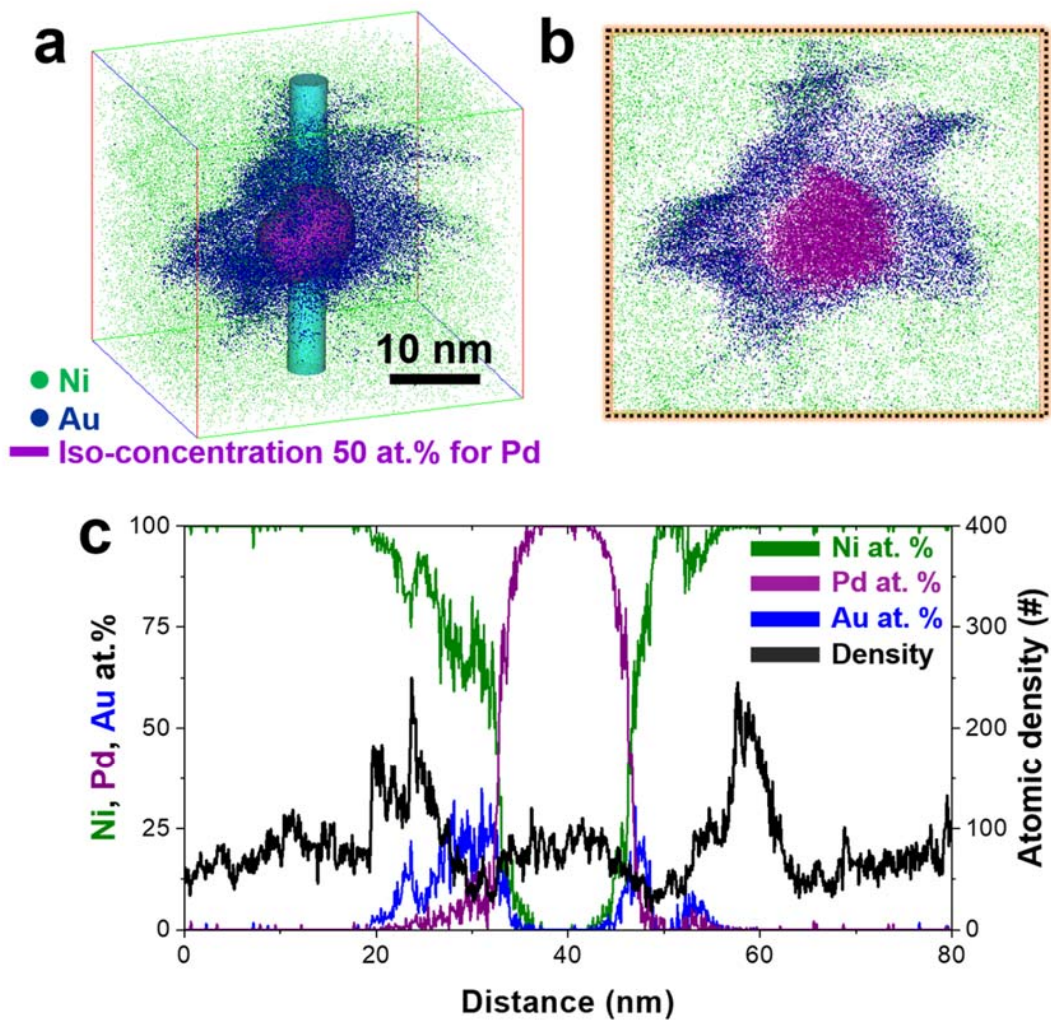

**Fig. 4.** (a) Reconstructed 3D atom map of Pd@Au core-shell nanoparticle embedded in Ni. The detected nanoparticle is highlighted by 50 at. % Pd iso-concentration surfaces. (b) 5 nm thin slice in x-direction sectioned from Fig. 4a. Green, dark-blue, and purple dots mark the reconstructed positions of Ni, Au, and Pd, respectively. (c) 1D concentration and atomic density (black) profiles measured vertically along the Pd@Au nanoparticle in Ni. Green, purple, and blue lines represent atomic concentrations for Ni, Pd, and Au respectively.

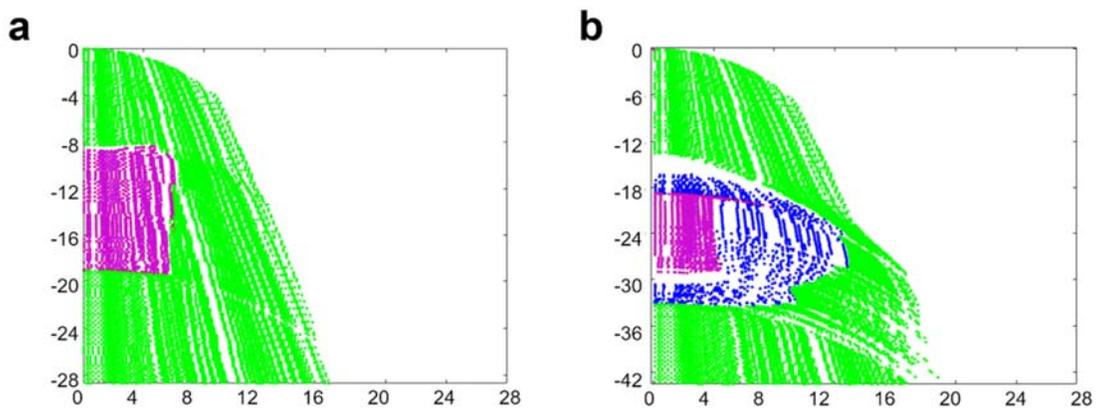

**Fig. 5.** Reconstructed atom maps acquired from field evaporation simulations of specimens containing (a) a square Pd nanoparticle (purple) and (b) a Pd@Au core-shell (purple@blue) nanoparticle within a Ni matrix (green). Evaporation fields assumed for Au, Pd, and Ni were 53, 37, and 35 V nm$^{-1}$, respectively, based on the image hump model [48] (scale units: nm).

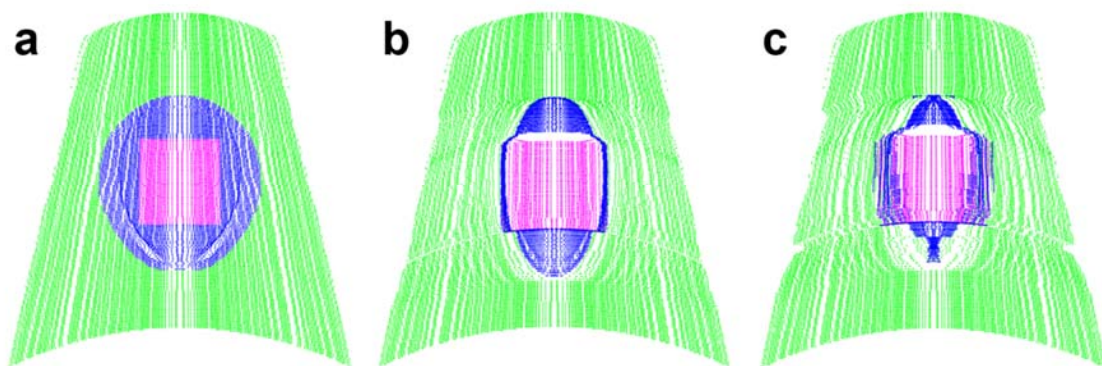

**Fig. 6.** Reconstructed atom maps acquired from field evaporation simulations of a Pd@Au core-shell (purple@blue) particle within a Ni matrix (green) with different evaporation fields of Au: (a) 35, (b) 28, and (c) 25 V nm$^{-1}$. Evaporation fields of Pd and Ni were fixed at 37 V nm$^{-1}$ and 35 V nm$^{-1}$, respectively.

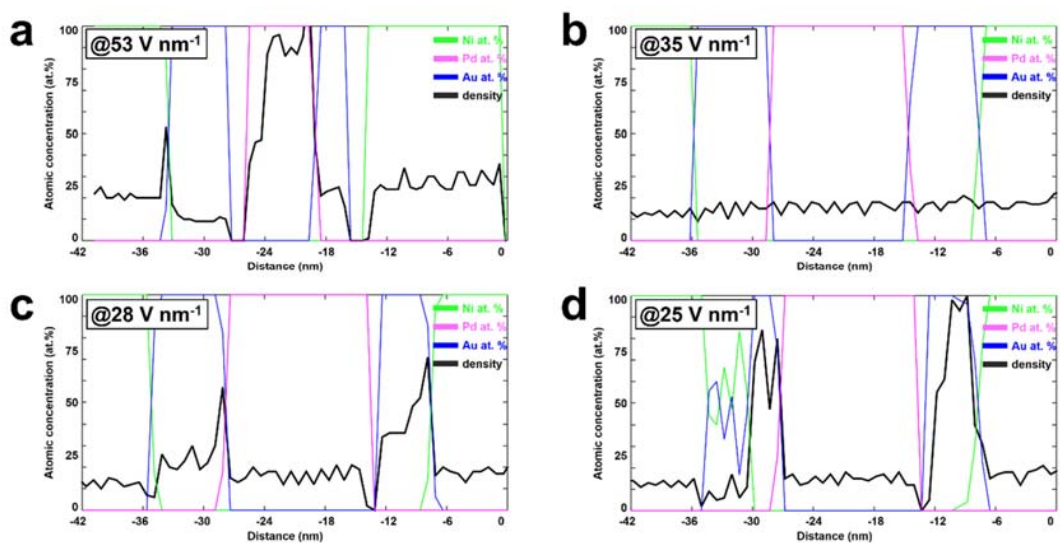

**Fig. 7.** 1D concentration and atomic density profiles measured vertically across the Pd@Au nanoparticles in the simulated datasets in Fig. 5 (a) and 6 (a) to (c). Light green, purple, and blue lines represent atomic concentrations for Ni, Pd, and Au respectively. The black lines represent atomic density profiles. Evaporation fields of Au were set as (a) 53, (b) 35, (c) 28, and (d) 25 V nm$^{-1}$.

**Table 1.** Evaporation fields (F$_e$) and field ratios (ε) of elements, as predicted by the image hump model [48], and input parameters of relative evaporation fields for field evaporation and ion trajectory simulation for Fig. 5.

|  | Evaporation field F$_e$ (V nm$^{-1}$) (according to Ref. [48]) | Relative evaporation field | Relative evaporation field used in simulation |
|---|---|---|---|
| Ni | 35 |  |  |
| Pd | 37 |  |  |
| Au | 53 |  |  |
| ε (Pd/Ni) |  | 1.06 |  |
| ε (Au/Ni) |  | 1.51 |  |
| Ni |  |  | 1.0 |
| Pd |  |  | 1.1 |
| Au |  |  | 1.5 |

# Supporting Information

**Characterization of Pd and Pd@Au core-shell nanoparticles using atom probe tomography and field evaporation simulation**

Se-Ho Kim[a,b], Kyuseon Jang[a], Phil Woong Kang[c], Jae-Pyoung Ahn[d], Jae-Bok Seol[e], Chang-Min Kwak[f], Constantinos Hatzoglou[g], François Vurpillot[g], Pyuck-Pa Choi[a*]

[a]Department of Materials Science and Engineering, Korea Advanced Institute of Science and Technology (KAIST) 291 Daehak-ro, Yuseong-gu, Daejeon 34141, Republic of Korea

[b]Max-Planck-Institut für Eisenforschung GmbH, Max-Planck-Straße 1, 40237, Düsseldorf, Germany

[c]Department of Chemical and Biomolecular Engineering, Korea Advanced Institute of Science and Technology (KAIST) 291 Daehak-ro, Yuseong-gu, Daejeon 34141, Republic of Korea

[d]Advanced Analysis Center, Korea Institute of Science and Technology (KIST), Seoul 136-791, Republic of Korea

[e]National Institute for Nanomaterials Technology (NINT), POSTECH, Pohang 790-784, Republic of Korea

[f]Department of Materials Science and Engineering, POSTECH, Pohang 790-784, South Korea

[g]Normandie Université, UNIROUEN, INSA Rouen, CNRS, Groupe de Physique des Matériaux, 76000 Rouen, France

*Corresponding author, Email: p.choi@kaist.ac.kr

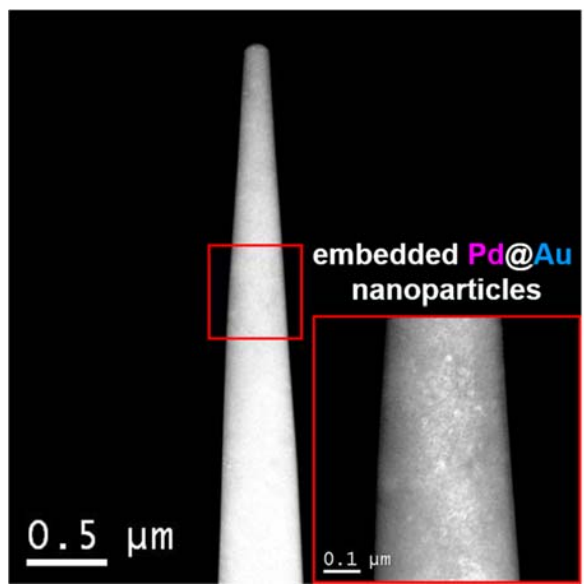

**Fig. S1.** HAADF-STEM image of as-sharpened APT specimen. The bright contrast shows that the nanoparticles are encapsulated well in Ni.